
%
%
\documentstyle{amsppt}
\magnification=\magstep1
\NoBlackBoxes
\define \cn{\text{cn}}
\define \sn{\text{sn}}
\define \dn{\text{dn}}
\topmatter
\title{ SEPARATION OF VARIABLES AND THE XXZ GAUDIN MAGNET${}^{\,1}$ }
\endtitle
\author
E.G. KALNINS, V.B. KUZNETSOV and WILLARD MILLER, Jr.
\endauthor
\affil
Department of Mathematics and Statistics,\\
University of Waikato,\\
Hamilton, New Zealand\\
\\
Department of Mathematics ${}^{\dag\,\S}$\\
University of Amsterdam, Plantage Muidergracht 24,\\
1018 TV Amsterdam, The Netherlands \\
\\
School of Mathematics ${}^{\ddag}$\\
and Institute for Mathematics and its Applications,\\
University of Minnesota,\\
Minneapolis, Minnesota  55455, USA. \\
\endaffil
\thanks{${}^1$ Will appear as Preprint in Mathematical preprint series,
University of Amsterdam, in January 1995; hep-th/9412190 }
\endthanks
\thanks{$\dag$ Work supported by the Netherlands Organisation
for Scientific
Research (NWO)}
\endthanks
\thanks{$\ddag$ Work supported in part by the National Science
Foundation under grant DMS 94--00533}
\endthanks
\thanks{$\S$ On leave from Department of Mathematical and
Computational Physics,
Institute of Physics, St. Petersburg University,
St. Petersburg 198904,
Russia}
\endthanks
\abstract{In this work we generalise  previous results
connecting  (rational) Gaudin magnet models and classical separation of
variables. It is shown that the connection persists for the case of linear
$r$-matrix algebra which corresponds to the trigonometric $4\times 4$
$r$-matrix (of the XXZ type). We  comment also on the
corresponding problem for the elliptic (XYZ) $r$-matrix. A prescription
for obtaining integrable systems associated with multiple poles of an
$L$-operator is  given.}
\endabstract
\endtopmatter
\document
%
%
\subheading{1. Introduction}
Separation of variables for the Hamilton-Jacobi and Schr\"odinger
equations have long been known as methods for explicit solution of these
equations in appropriate circumstances. The technical requirements for this
method of solution have quite fully developed in recent years
(see \cite{1--8}). In particular the relationship between the separable
systems and the Gaudin magnet \cite{4,9} integrable systems models has been
established via $r$-matrix algebra methods, where the $r$-matrix
corresponds to the rational or so called XXX case, \cite{4--8}. This
relationship works very clearly with  separable coordinate systems on spaces
of constant curvature. The question we answer here is how  these notions
can be extended to include the so-called trigonometric $r$-matrix algebra
in the XXZ case.  To do this let us recall the fundamental ideas of the
$r$-matrix formalism (see \cite{10,11}
and references in there). For a classical
mechanical system the basic (linear) $r$-matrix algebra is
$$
\{L(u)\otimes I,I\otimes L(v)\}=
\left[r(u-v),\ L(u)\otimes I+I\otimes L(v)\right]\,,\tag1.1
$$
where $\{\cdot,\cdot\}$ is the Poisson bracket and $[\cdot,\cdot]$ the
matrix commutator bracket. The operator $L(u)$ is  taken to be the
$2\times 2$ matrix
$$
L(u)_{11}= -L(u)_{22} = A(u),\qquad L(u)_{12} = B(u),\qquad
L(u)_{21}=C(u),
$$
and $r(u)$ is a suitable $4\times 4$ matrix of scalars solving the classical
Yang-Baxter equation \cite{10,11}; $u$ being arbitrary constant
is called the spectral parameter. In the case of the XXZ $r$-matrix
algebra the non zero elements of $r$ can be taken to be
$$
r(u)_{11}=r(u)_{44}=\coth(u),\qquad
r(u)_{23}=r(u)_{32}=\frac{1}{\sinh(u)}.\tag1.2
$$
In component form, the $r$-matrix algebra relations are
$$\align
\{A(u),A(v)\}&=\{B(u),B(v)\}=\{C(u),C(v)\}=0\,,\\
\{A(u),B(v)\}&=\frac{1}{\sinh(u-v)}\left(\cosh(u-v)B(v)-B(u)\right),\tag1.3\\
\{A(u),C(v)\}&=\frac{1}{\sinh(u-v)}\left(-\cosh(u-v)C(v)+C(u)\right),\\
\{B(u),C(v)\}&=\frac{-2}{\sinh(u-v)}\left(A(u)-A(v)\right).
\endalign$$

If we now make the ansatz $A(u)=\coth(u)S_3,\ B(u)=(1/\sinh(u))S_+$ and
$C(u)=(1/\sinh(u))S_-$ these relations imply
$$
\{S_3,S_\pm \}=\pm S_\pm \, ,\qquad \{S_+,S_-\}=2S_3\,. \tag1.4
$$

To relate this observation to the separation of variables methods,
we form the $L(u)$ operator with elements
$$\align
B(u)=&\sum _{\alpha=1}^n \frac{1}{\sinh(u-e_\alpha )}\,S_{+\alpha }\,,
\qquad
C(u)=\sum _{\alpha=1}^n\frac{1}{\sinh(u-e_\alpha )}\,S_{-\alpha }\,,\\
A(u)=&\sum _{\alpha=1}^n \coth(u-e_\alpha )\,S_{3\alpha }\,, \tag1.5
\endalign$$
where
$$
\{S_{3\alpha },S_{\pm \beta }\}=\pm \delta _{\alpha \beta }S_{\pm \alpha }\,,
\qquad
\{S_{+\alpha },S_{-\beta }\}=2\delta _{\alpha \beta }S_{3\alpha }\,.\tag1.6
$$

The $r$-matrix algebra relations, (1.1) or (1.3), imply
$$
\{\det L(u),\det L(v)\}=0\,, \tag1.7
$$
i.e., that $\det L(u)$ is a generating function of the constants of the motion.
In particular we have
$$
-\det L(u)=A^2(u)+B(u)C(u)
=\sum _{\alpha=1}^n \left[\frac{C_\alpha}{\sinh^2(u-e_\alpha )} +
H_\alpha \coth(u-e_\alpha
)\right]+H^2_0, \tag1.8
$$
where $C_\alpha =(S_{3\alpha })^2+S_{+\alpha }S_{-\alpha }$ are the Casimir
elements of the algebra generated by elements $S_{+\alpha } ,\ S_{-\alpha }$
and
$S_{3\alpha }$.  Furthermore
$H_0 =\sum _\alpha S_{3\alpha }$
and
$$
H_\alpha =\sum _{\beta \neq \alpha }\left(2S_{3\alpha }S_{3\beta }
\coth(e_\alpha -
e_\beta )+\frac{1}{\sinh(e_\alpha -e_\beta )}(S_{+\alpha }S_{-\beta }+
S_{+\beta }S_{-
\alpha })\right). \tag1.9
$$
With the following realization of the algebra in terms of the canonical
coordinates $x_\alpha$ and $p_\alpha$, $\{p_\alpha,x_\beta\}=
-\delta_{\alpha\beta}$ :
$$
S_{+\alpha }=\frac{i}{2}x^2_\alpha\,, \qquad
S_{-\alpha }=\frac{i}{2}p^2_\alpha\,, \qquad
S_{3\alpha }=-\frac{1}{2}x_\alpha p_\alpha\,, \tag1.10
$$
the constants (1.9) have the form
$$
H_\alpha =\frac14\sum _{\beta \neq \alpha }\frac{-1}{\sinh(e_\alpha -
e_\beta )}\left(x^2_
\alpha p^2_\beta +x^2_\beta p^2_\alpha -2x_\alpha x_\beta p_\alpha p_\beta
\cosh(e_\alpha -e_\beta )\right),  \tag1.12
$$
and $H_0=-\frac12\sum _\alpha x_\alpha p_\alpha\,$.
Notice that all $C_\alpha=0$ in such a representation.
%
%
\subheading{2. Variable Separation for the XXZ Magnet}
Proceeding as in \cite{4,7,8}, we choose separable coordinates such that
$B(u)=0$, i.e.,  $u=u_j ,\ j=1,\ldots,n-1$. This implies
$\sum _\alpha x^2_\alpha /$sinh$(u-e_\alpha )=0$ for $u=u_1,\ldots,u_{n-1}$,
which in turn implies that we choose coordinates according to
$$
x^2_\alpha =e^{u_n}\,\frac{\Pi ^{n-1}_{j =1}\sinh(u_j -e_\alpha )}{\Pi _{
\beta \neq \alpha }\sinh(e_\beta -e_\alpha )}\,, \tag2.1
$$
motivated by the general formula
$$
\sum ^{n}_{\alpha =1}\frac{x^2_\alpha}{\sinh(u-e_\alpha )}=e^{u_n}
\,\frac{\Pi ^{n-1}_
{j =1}\sinh(u-u_j )}{\Pi ^n_{\beta =1}\sinh(u-e_\beta)}\,. \tag2.2
$$

For each $u_j$ we can define the canonically conjugate coordinate $v_j$ as
follows:
$$
v_j=A(u_j)=-\frac{1}{2}\sum ^n_{\alpha =1}\coth(u_j-e_\alpha )\,x_\alpha
p_\alpha\,,\quad
1\le j\le n-1\,,\;\quad v_n=H_0\,.  \tag2.3
$$
The coordinates $u_i,\ v_j$ ($i,j=1,\ldots,n$)
satisfy the canonical bracket relations
$$
\{u_i,u_j\}=0\,,\quad \{v_i,u_j\}=\delta _{ij}\,,
\quad \{v_i,v_j\}=0\,. \tag2.4
$$

The changing of variables $x_\alpha,p_\alpha $ for the new variables
$u_i, v_i$ is
the procedure of separation of variables. The matrix elements of the
$L$-operator can be expressed in terms of these variables  according to the
formulas
$$\align
A(u)=&iB(u)\left[
2\cosh\left(
u+\Sigma^{n-1}_{j=1}u_j-\Sigma^{n}_{\alpha =1}e_\alpha \right)\;v_n
 \right.\\
&\left.
\quad\qquad
\qquad+\sum ^{n-1}_{j=1}\frac{-2v_j}{\sinh(u-u_j)}\frac{\Pi ^{n}_{\alpha =1}
\sinh(u_j-e_\alpha )}
{\Pi _{k\neq j}\sinh(u_j-u_k)}\right]e^{-u_n}.\tag2.5
\endalign$$
The entry $C(u)$ can be computed by using the formula
$$\align
p_\alpha =
&x_\alpha e^{-u_n} \left[
2\cosh\left(\Sigma^{n-1}_{j=1}u_j-\Sigma_{\gamma
\neq \alpha }e_
\gamma \right)\;v_n  \right.\\
&\left.\quad\qquad\qquad
+\sum ^{n-1}_{j=1}\frac{-2v_j}{\sinh(e_\alpha -u_j)}
\frac{\Pi ^{n}_{\alpha=1}\sinh(u_j-e_\alpha)}
{\Pi _{k\neq j}\sinh(u_j-u_k)}\right].\tag2.6
\endalign$$
This gives the relation between the coordinates $x_\alpha ,p_\alpha $ and
$u_i,v_i$ where $v_{n}=H_0$.

The equation for the eigenvalue curve $\Gamma:\;\det(L(u)-\lambda I)=0$, has
the form
$$\lambda ^2-A(u)^2-B(u)C(u)=0\,.$$
If we put $u=u_j,\;j=1,\ldots,n-1$
into this equation then $\lambda =\pm v_j$. Thus variables
$u_j$ and $v_j$ ($j=1,\ldots,n-1$) lie on the curve $\Gamma $:
$$v^2_j- \sum ^n_{\alpha =1} H_\alpha \coth(u_j-e_\alpha)-H_0^2\equiv v^2_j+
\det L(u_j)
=0\,.\tag2.7
$$
Equations (2.7) are the separation equations for the
degrees of freedom connected with the values of the integrals $H_\alpha$.
(Note that $\sum_{\alpha=1}^n H_\alpha=0$.)

For  illustrative reasons it is more transparent
to use the variables $A_i=e^{e_i}$ and $U_i=e^{u_i}$. Then many
of the expressions given have algebraic form. For example, the nearest
object we have to a Hamiltonian in the case of XXZ $r$-matrix algebra is
$H=\sum ^n_{i=1}A^2_i H_i$ which has the form
$$\align
H&=\sum ^{n-1}_{i=1}
e^{-\sum_{j\neq i}u_j}\left(e^{-2u_i}v^2_i+H^2_0\right)
{\Pi ^{n}_{\alpha =1}\hbox{sinh}(u_i-e_\alpha )\over
\Pi _{j
\neq i}\hbox{sinh}(u_i-u_j)}\\
&=(4\Pi ^n_{j=1}A_j)^{-1}(-1)^n\sum ^{n-1}_{i=1}
\left(p^2_{U_i}+H^2_0\right)
{\Pi ^{n}_{k=1}(A^2_k-U^2_i)
\over U_i^2\Pi _{j\neq i}(U^2_i-U^2_j)}\,. \tag2.8
\endalign$$
Note that
$$ e^{u_n}=\frac{U_1\cdots U_{n-1}}{A_1\cdots A_n}\sum_{\alpha=1}^n
A_\alpha x_\alpha^2\,.
$$

If we adopt the standard procedure and write $p_{U_i}=
\partial W/\partial U_i$ then the equation $H=E$ admits separation of variables
via the usual ansatz $W=\sum ^{n-1}_{i=1}W_i(U_i).$
The separation equations can be written in the alternate form
$$
\left[\Pi ^{n}_{k=1}(U^2_j-A^2_k)\right]
\left(\frac{\partial W_j}{\partial {U_j}}\right)^2=
H_0U^{2n-2}_j+H_0(-1)^n \left(\Pi_kA_k^2\right)U_j^{-2}+
\sum ^{
n-1}_{k=1}\lambda_kU^{2k-2}_j,
$$
where the $\lambda_k$ are related to the $H_j$ via
$$
P_{2n-4}(u)\equiv 2u^{-2}\left[\Pi_{k=1}^n(u^2-A_k^2)\right]\sum_{i=1}^n
\frac{H_iA_i^2}{u^2-A_i^2}\,,
$$
$$
\lambda_k=H_0\sum_{\alpha_1<\ldots<\alpha_{n-k}}(-1)^{n-k}
A_{\alpha_1}^2\cdots A_{\alpha_{n-k}}^2+\frac{1}{(2k-2)!}
P_{2n-4}^{(2k-2)}(0)\,,
$$
$$
\lambda_{n-1}=E-H_0\sum_{k=1}^nA_k^2\,.
$$
\medskip

{}From what has been developed so far we see that separation of variables goes
through for  XXZ $r$-matrix algebras constructed in this way. In the
previous article \cite{8} for the case of spaces of constant curvature  we
essentially have the rational $r$-matrix algebra and it is possible to
formulate using well defined limiting procedures the cases of integrable
systems for which some of the $e_i$ parameters are equal. What was also
established previously was the construction of integrable systems
given on the algebra with commutation relations
$$
\{(Z^J_j)_\ell
,(Z^{J'}_k)_m\}=-\delta _{JJ'}(Z^J_{j+k-N_J})_s\epsilon_{\ell ms}, \tag2.9
$$
where $\ell,m,s=1,2,3,\ 0<j<N_J,\ 0<k<N_{J'},\  0<J\le p$
and $\epsilon _{\ell ms}$
is the usual totally antisymmetric tensor, and the vector $Z_j^J$ has the form
$$\align
&Z^J_j=\pmatrix
{1\over 4} \sum _i(p^J_ip^J_{j+1-i}+x^J_ix^J_{j+1-i})\cr {i\over 4}
\sum _i(p^J_ip^J_{j+1-i}-x^J_ix^J_{j+1-i})\cr
{i\over 2} \sum _ip^J_ix^J_{j+1-i}\endpmatrix
 \tag2.10
\endalign$$
in the coordinate representation. Indeed, if we adopt the limiting procedure
$$\align
A^J_j&\rightarrow  A^J_1+{}^J\epsilon ^1_{j-1} \,,
\quad j=1,\ldots,N_J \,,\quad J=1,\ldots,p\,,\\
p^J_j&\rightarrow \sqrt{a^J_j}
\left(p^J_1+\sum ^{N_{J}}_{i=2}{}^J\epsilon ^{i-1}_{
j+1-i} \,p^J_i\right),\\
x^J_j&\rightarrow \sqrt{a^J_j}
\left(x^J_1+\sum ^{N_{J}}_{i=2}{}^J\epsilon ^{i-1}_{
j+1-i} \,x^J_i\right), \tag2.11
\endalign$$
where
$$
{}^J\epsilon ^{i-1}_{j+1-i}=
\Pi ^i_{\ell =2}({}^J\epsilon ^1_{j-1}-{}^J\epsilon ^1_{\ell-2} )\,,\qquad
a^J_j=\frac{1}{
\Pi _{k\neq j}({}^J\epsilon ^1_{j-1}-{}^J\epsilon ^1_{k-1} )}\,,
$$
and $N_1+...+N_p=n+1$, then the Hamiltonian $H$ has the form
$$
H=(\Pi^{n+1}_{j=1}A_j)^{-1}
\sum ^n_{i=1}\left(p^2_{U_i}+H^2_0\right)
{\Pi ^p_{k=1}(A^2_k-U^2_i)^{N_k}\over U_i^2\Pi _{j%
\neq i}(U^2_i-U^2_j)}   \tag2.12
$$
with obvious separation equations. (We require ${}^J\epsilon_0^i=0$ and take
the limit as the ${}^J\epsilon_h^1\rightarrow 0$ for $h=1,\ldots,N_J-1$,
see \cite{8}.) The generating function
for the constants can
be derived by applying these procedures to $\det L(u)$. We will, however,
adopt a
different and more general strategy. If we leave the matrix elements of $L(u)$
in the form (1.5) and subject the resulting expression for $\det L(U)$
$$\align
-\frac{1}{2}\det L (U)&=\frac{1}{2}\left(\sum_{\alpha=1}^n
S_{3\alpha}\right)^2
+2U^2\left[ \sum ^2_{i=1} \left(\sum ^n_{\alpha =1}\frac{A_\alpha
S_{i\alpha }}{
U^2-A^2_\alpha} \right)^2 \right.\\
&\left.+\sum ^n_{\alpha =1}\frac{S_{3\alpha }}{
U^2-A^2_\alpha}\left(-\sum ^n_{\alpha =1}S_{3\alpha }+
U^2\sum ^n_{\alpha =1}\frac{S
_{3\alpha }}{U^2-A^2_\alpha}\right)\right],  \tag2.13
\endalign$$
where $U=e^u$, and $S_{\pm\alpha}=S_{1\alpha}\pm iS_{2\alpha}$, to the
transformations
$$\align
&A^J_j\rightarrow  A^J_1+{}^J\epsilon ^1_{j-1} \,,\qquad j=1,\ldots,N_J \,,
\quad J=1,\ldots,p\,,\\
&{}^JS_1\delta _{k0} + \sum ^{N_J}_{j=2}({}^J\epsilon _{j-1}^1)^k({}^JS_j)=
Z^J_{N_J-k} \,,\quad  k=0,\ldots,N_J\,,
\endalign$$
then we arrive at a general expression for the generating function $\det L(U)$.
The constants of the motion are obtained by the usual means of expanding
the expression following from (2.13)
in partial fractions and reading off the independant
components. In the case of degenerate roots the expression can be readily
modified. Accordingly we have
$$\align
-\frac{1}{2}\det L(U)&=\frac{1}{2}\left(\sum_{J=1}^p
(Z^J_{N_J})_3\right)^2\\
&+2U^2\left[\sum ^2_{i=1}\left(\sum ^p_{J=1}
\sum ^{N_J-1}_{j=0}\frac{1}{j!}\left(
\left(\frac{\partial}{\partial A_j}\right)^j\frac{A_j}{
U^2-A^2_j}\right) (Z^J_{N_J-j})_i\right)^2\right.\\
&+\sum ^p_{J=1}
\sum ^{N_J-1}_{j=0}\frac{1}{j!}\left(
\left(\frac{\partial}{\partial A_j}\right)^j\frac{1}
{U^2-A^2_j}\right)
(Z^J_{N_J-j})_3\times\tag2.14\\
&\left.\times\sum ^p_{J=1}\left(-(Z^J_{N_J})_3+U^2
\sum ^{N_J-1}_{j=0}\frac{1}{j!}\left(
\left(\frac{\partial}{\partial A_j}\right)^j\frac{1}
{U^2-A^2_j}\right)
(Z^J_{N_J-j})_3\right)\right].
\endalign$$

{}From this expression constants of the motion can be deduced just as before.
The separation of variables proceeds as usual in the case of the choice of
coordinates as given in \cite{8}. The expressions for the coordinates
corresponding
to multiple roots with signature $N_1,N_2,...,N_p$ can be obtained from the
generic case by the limiting procedures already outlined. In rational form the
generic coordinates are
$$
x^2_i =[\Pi ^{n}_{j=1}(A^2_j
-U^2_i)][\Pi ^{n-1}_{s=1}U_s]
\frac{\Pi _{k\neq i}A_k}{\Pi _{\ell\neq i}(A^2_\ell -A^2_i)}\,.\tag2.15
$$
For the case of signature $N_1,N_2,...,N_p$ the coordinates are given by the
relations
$$\align
\sum ^j_{i=1}x^J_ix^J_{j+1-i}
&=[\Pi ^n_{k=1}U_k]\left\{\sum ^{j-2}_{r=1}(N_J-r-3)
[\Pi ^{r-2}_{q=0}(N_J+q-2)]\times\right.\\
&\left.\times\frac{[\Pi _
{L\neq J}(A^L_1)^{N_L}]}{(2^{N_J})A^J_1r!}\left(\frac{\partial}
{\partial A^J_1}\right)^r\frac{\Pi ^n_{i=1}((A^J_1)^2-U^2_i)
}{A^J_1 \Pi _{L\neq J}(2A^L_1)^{N_L}}\right\}.
\endalign$$
This  gives a complete description of the separation of variables procedure
for the signature  $N_1,N_2,...,N_p$ case. We illustrate these ideas with
two examples.

{\bf A. The case of signature 2,1 and dimension 3.}
In this case the generating function assumes the form
$$\align
\det L(u)&=\frac{1}{\sinh^2(u-e_1)}
\left\{\frac12(Z^1_1.Z^1_1+Z^1_2.Z^1_2)-
\frac{1}{\sinh(e_1-e_3)}
\left((Z^1_1)_2(Z^2_1)_2\right.\right.\\
&\left.+(Z^1_1)_1(Z^2_1)_1\right)+\cosh(e_1-e_3)(Z^1_1)
_3(Z^2_1)_3\}\\
&+\frac{ Z^1_1.Z^1_1\coth(u-e_1)}{2\sinh^4(u-e_1)}
\left\{\frac{1}{\sinh(e_1-e_3)}\left((Z^1_2)_1(Z^2_1)_1+(Z^1_2)_2(Z^2_1)_2
\right)\right.\\
&+\cosh(e_1-e_3)(Z^1_2)_3(Z^2_1)_3-\frac{1}{\sinh^2(e_1-e_3)}
\left((Z^1_1)_3(Z^2_1)_3\right.\\
&\left.\left.\left.
+\cosh(e_1-e_3)((Z^1_1)_2(Z^2_1)_2+(Z^1_1)_1(Z^2_1)_1\right)\right)\right\}\\
&-\coth(u-e_3)
\left\{\frac{1}{\sinh(e_1-e_3)}\left(-\left((Z^2_1)_2(Z^1_2)_2+
(Z^2_1)_1(Z^1_2)_1\right)\right.\right.\\
&\left.+\cosh(e_1-e_3)(Z^2_1)_3(Z^1_2)_3\right)
+\frac{1}{\sinh^2(e_1-e_3)}\left((Z^2_1)_3(Z^1_1)_3\right.\\
&\left.\left.+\cosh(e_1-e_3)\left((Z^2_1)_1(Z^1_2)_1+(Z^2_1
)_2(Z^1_2)_2\right)\right)\right\}.
\endalign$$

The constants of the motion can be deduced from the coefficents of independent
functions of $u$. In the coordinate representation these constants
have the form
$$\align
H_1&=x^2_1p^2_2+p^2_1x^2_2-2x_1x_2p_1p_2+\frac{1}
{\sinh(e_1-e_3)}\left(x^2_1p^2_3+x^2_3p^2_1-2x_1x_3p_1p_3\right),\\
H_2&=\frac{2}{\sinh(e_1-e_3)}\left(x_1x_2p^2_3+p_1p_2x^2_3-
(x_1x_3p_2p_3+p_1p_3x_2x_3)\cosh(e_1-e_3)\right)\\
&-\frac{1}{\sinh^2(e_1-e_3)}\left(x_1p_3+p_1x_3\right)^2,
\endalign$$
\medskip
\noindent
where we have used the notation $x_1=x^1_1\,,
\;x_2=x^1_2$ and $x_3=x^2_1\,$, with
similar relations for the $p_i$'s. The coordinates are given by the formulas
$$\align
x^2_1& = -\frac{\sinh(u_1-e_1)\sinh(u_2-e_1)}{\sinh(e_1-e_3)}\,,\\
2x_1x_2&=-\frac{\sinh(u_1-e_1)\sinh(u_2-e_1)\cosh(e_1-e_3)}
{\sinh^2(e_1-e_3)}-\frac{1}{\sinh(e_1-e_3)}\times\\
&\times\left(\sinh(u_1-e_1)\cosh(u_2-e_1)
+\sinh(u_1-e_3)\cosh(u_2-e_3)\right),\\
&x^2_3 =\frac{\sinh(u_1-e_3)\sinh(u_2-e_3)}{\sinh^2(e_1-e_3)}\,.
\endalign$$

{\bf B. The Case of Signature 3 and Dimension 3.}
The generating function has the form
$$\align
L(U)&=\frac{1}{(U^2-A^2_1)^6}32A^8_1 Z_1.Z_2\\
&+\frac{1}{(U^2-A^2_1)^5}16A^2_1\left[5Z_1.Z_1+(Z_1)_2(Z_2)_2+
2A_1\left((Z_1)_3(Z_2)_3+(Z_1)_1(Z_2)_1\right)\right]\\
&+\frac{1}{(U^2-A^2_1)^4}2A^4_1\left[33Z_1.Z_1+36Z_1.Z_2+8A^2_1Z_1.Z_3
+4A^2_1Z_2.Z_3\right]\\
&+\frac{1}{(U^2-A^2_1)^3}2A^2_1\left[26A_1Z_1.Z_2+8A^2_1Z_2.Z_2\right.\\
&\left.+ 14A^2_1Z_1.Z_3+Z_1.Z_1+4A^3_1Z_3Z_2+(Z_1)^2_3\right]\\
&+\frac{1}{(U^2-A^2_1)^2}2\left[A^4_1Z_3.Z_3+6A^3_1Z_2.Z_3\right.\\
&+A^2_1\left(5\left((Z_2)^2_2+(Z_2)^2_1\right)+6(Z_3)_2(Z_1)_1+4(Z_2)^2_3
\right)\\
&\left.+A_1\left(6(Z_2)_2(Z_1)_2+(Z_2)_1(Z_1)_1+4(Z_2)_3(Z_1)_3+(Z_1)^2
\right)\right]\\
&+\frac{1}{U^2-A^2_1}2\left[A^2_1Z_3.Z_3+2A_1Z_2.Z_3+(Z_2)^2_2+(Z_2)^2_1+
(Z_1)_3(Z_3)_3\right].
\endalign$$
The coordinates are given by
$$\align
x^2_1&=\frac14U_1U_2-\frac{A^2_1}{4}\left(\frac{U_1}{U_2}
+\frac{U_2}{U_1}\right)+\frac{A^4_1}{4U_1U_2}\,,\\
2x_1x_2&=-\frac{3}{8A_1}U_1U_2-\frac{A_1}{8}\left(\frac{U_1}{U_2}
+\frac{U_2}{U_1}\right)+\frac{5A^3_1}{8U_1U_2}\,,\\
2x_1x_3+x^2_2&=\frac{3U_1U_2}{8A^2_1}+\frac18\left(\frac{U_1}{U_2}
+\frac{U_2}{U_1}\right)+\frac{3A^2_1}{8U_1U_2}\,.
\endalign$$
%
%
\subheading{3. The XYZ Magnet}
These methods can be extended to the case of elliptic or XYZ $r$-matrix
algebras. The only difference is that in this case a solution of the
problem via
separation of variables is not yet known \footnote[1]{See \cite{12}
where the variable separation has been done
for the periodic classical XYZ-chain from which the system in question
can be obtained through the limit.}
but the coalescing of indices goes
through just as before. Indeed, the operator
$L(u)$ can be taken just as in (1.1). The non zero elements of the
$r$-matrix in
this case are
$$\align
r(u)_{11}&=r(u)_{44}=\frac{\text{cn}(u)}{\sn(u)},\qquad
r(u)_{14}=r(u)_{41}= \frac{1-\dn(u)}{2\sn(u)}\,,\\
r(u)_{23}&=r(u)_{32}=\frac{1+\dn(u)}{2\sn(u)}\,. \tag3.1
\endalign$$

We now make the ansatz
$$\align
A(u)&=\frac{\cn(u)}{\sn(u)}S_3 \,,\quad
B(u)=\frac{1}{2\sn(u)}\left[(1+\dn(u))S_-+(1-\dn(u))S_+\right]\,,\\
C(u)&=\frac{1}{2\sn(u)}\left[(1+\dn(u))S_++(1-\dn(u))S_-\right]\,. \tag3.2
\endalign$$
Here the $S_\pm $, $S_3$ obey the same  commutation relations as (1.4). We
choose the $L(u)$ operator to be
$$\align
A(u)&=\sum_{\alpha=1}^n \frac{\cn(u-e_\alpha )}{\sn(u-e_\alpha )}
S_{3\alpha }\,,\tag3.3\\
B(u)&=\sum_{\alpha=1}^n \frac{1}{2\sn(u-e_\alpha )}
\left[(1+\dn(u-e_\alpha ))S_{-\alpha}
+(1-\dn(u-e_\alpha ))S_{+\alpha}\right]\,,\\
C(u)&=\sum_{\alpha=1}^n \frac{1}{2\sn(u-e_\alpha )}
\left[(1+\dn(u-e_\alpha ))S_{+\alpha}
+(1-\dn(u-e_\alpha ))S_{-\alpha }\right]\,.
\endalign$$

The determinant of $L(u)$ is once again a generator of the constants of the
motion. It has the form
$$
\det L(u)=\sum^n_{\alpha =1}H_\alpha E(u-e_\alpha +iK')+H_0  \tag3.4
$$
where
$$\align
H_\alpha& =2k^2\sum _{\beta \neq \alpha }\frac{1}{\sn(e_\alpha -e_\beta )}
\left[S_{1\alpha }S_{1\beta }+\dn(e_\alpha -e_\beta )S_{2\alpha }S_{2\beta }
+\cn(e_\alpha -e_\beta )S_{3\alpha }S_{3\beta }\right]\,,\\
H_0&=2k^2{\sum_{\alpha,\beta}}\;{}^{'}\;\frac{E(e_\beta -e_\alpha )}
{\sn(e_\beta -e_\alpha)}
 [S_{1\alpha }S_{1\beta }+\dn(e_\alpha -e_\beta )S_{2\alpha }S_{2\beta }+
\cn(e_\alpha -e_\beta )S_{3\alpha }S_{3\beta }]\\
&-{\sum_{\alpha,\beta}}\;{}^{'}\;
[k^2\cn(e_\alpha -e_\beta )S_{2\alpha }S_{2\beta }+
\dn(e_\alpha -e_\beta )S_{3\alpha }S_{3\beta }]
-\sum^n_{\alpha =1}[k^2S^2_{2\alpha }+S^2_{3\alpha }]\,. \tag3.5
\endalign$$
Here $E(z) =\int ^z\dn^2(u)du$ is Jacobi's epsilon function.
The same is now true
as for the case of XXZ $r$-matrix algebras: if we subject the $e_\alpha$'s
and the $S_{i\beta }$'s to the transformations given by (2.11),
then we arrive at
the generating function for the constants of motion for a root structure
having
the signature $N_1,N_2,....,N_p$. The expression for this function is
$$
\det L(u)=\sum ^3_{k=1}\left(\sum ^{N_J-1}_{r=0}\left(\frac{\partial}
{\partial {}^Je_1}\right)^rf_k(u-{}^Je_1)(Z^J_{N_J-r})_k\right)^2,  \tag3.6
$$
where $f_1(z)=1/\sn(z)$, $f_2(z)=\dn(z)/\sn(z)$, and $f_3(z)=\cn(z)/\sn(z)$.

As an example, the generating function corresponding to signature 2,1 is
$$\align
\det L(u)&=H_1E(u-e_1+iK')+H_2E(u-e_3+iK')+H_3+\frac{1}{\sn^4(u-e_1)}
H_4\\
&+\frac{\cn(u-e_1)\dn(u-e_1)}{\sn^3(u-e_1)}H_5+\frac{1}{\sn^2(u-e_1)}H_6
+\frac{1}{\sn^2(u-e_3)}H_7,
\endalign$$
where
$$\align
H_1&=\frac{2}{k^2\sn(e_1-e_3)}\left[(Z^2_1)_1(Z^1_2)_1-k^2\cn(e_1-e_3)
(Z^2_1)_3(Z^1_2)_3\right.\\
&\left.-\frac{k^2\cn(e_1-e_3)\dn(e_1-e_3)}{\sn(e_1-e_3)}(Z^2_1)_1(Z^1_1)_1+
\frac{\dn(e_1-e_3)}{\sn(e_1-e_3)}(Z^2_1)_3(Z^1_1)_3\right],\\
H_2&=-H_1\,,\\
H_3&=\frac{E(e_1-e_3)}{k^2\sn(e_1-e_3)}H_1+
k^2\left((Z^1_1)^2_1-(Z^1_2)^2_2-(Z^2_1)^2_2\right)
-(Z^2_1)^2_2-(Z^2_1)^2_3\\
&-2k^2\sn(e_1-e_3)(Z^2_1)_1(Z^1_1)_1+2\dn(e_1-e_3)(Z^2_1)_3(Z^1_2)_3\,,\\
H_4&=Z^1_1.Z^1_1\,,\qquad H_5=2Z^1_1.Z^1_2\,,\\
H_6&=Z^1_2.Z^1_2-(Z^1_1)^2_1-(Z^1_1)^2_2-k^2(Z^1_1)^2_3+
\frac{2}{\sn(e_1-e_3)}(Z^2_1)_1(Z^1_1)_1\\
&-2\frac{\cn(e_1-e_3)}{\sn(e_1-e_3)}(Z^2_1)_3(Z^1_2)_3\,,\\
H_7&=Z^2_1.Z^2_1\,.
\endalign$$

We  note  that the ideas developed here also work in the case of
separation of variables for spaces of constant Riemannian curvature, as
developed in previous articles \cite{6--8}. Indeed, in that case the rational
$r$-matrix algebra is as before and the non zero elements of the $r$-matrix
are
$$
r(u)_{11}=r(u)_{44}=r(u)_{23}=r(u)_{32}=1. \tag3.7
$$
The generating function of the constants of the motion for signature
$N_1,....,N_p$ is then
$$
\det L(u)=\sum ^3_{k=1}\left(\sum ^p_{J=1}\sum ^{N_J-1}_{j=0}
\frac{(Z^J_{j+1})_k}{(u-{}^Je_1)^{N_J-j}}+\epsilon _k\right)^2. \tag3.8
$$
This is the generalisation of the generating function for separable coordinates
on spaces of constant curvature of dimension $n=\sum ^p_{j=1}N_J+1$. Indeed, if
we use the form (3.8) and if $\epsilon _k=0$ for $k=1,2,3$ then we have the
generating function on the sphere for generic ellipsoidal coordinates, and if
$\epsilon_1=-1/4$, $\,\epsilon_2=1/4$, $\,\epsilon_3=0$ then we have the
generating
function of ellipsoidal coordinates in $n$-dimensional Euclidean space. As an
example consider the system with signature 2,1. The generating function is then
$$\align
\det L(u)&=\frac{\frac{(Z^2_1).(Z^1_2)}{e_1-e_3}-\frac{(Z^1_1).(Z^1_2)}
{(e_1-e_3)^2}}{u-e_1}+
\frac{\frac{(Z^1_1).(Z^1_2)}{2}-\frac{(Z^2_1).(Z^1_1)}{e_1-e_3}}{(u-e_1)^2}\\
&+\frac{(Z^1_1).(Z^2_1)}{(u-e_1)^3} +\frac{(Z^1_1).(Z^1_1)}{(u-e_1)^4}
-\frac{\frac{(Z^2_1).(Z^1_2)}{e_1-e_3}-\frac{(Z^1_1).(Z^1_2)}
{(e_1-e_3)^2}}{u-e_3}+
\frac{(Z^2_1).(Z^2_1)}{(u-e_3)^2}\,.
\endalign$$
                             
\enddocument